\documentclass[final,5p,times,twocolumn]{elsarticle}
\usepackage{graphicx}
\usepackage{amssymb}
\usepackage{amsmath}

\begin{document}

\begin{frontmatter}
\title{Ferromagnetism, spiral magnetic structures and phase separation in the two--dimensional Hubbard model}

\author{P. A. Igoshev\corref{cor1}}
\ead{Igoshev\underline{ }PA@imp.uran.ru}
\author{A. V. Zarubin}
\author{A. A. Katanin}
\author{V. Yu. Irkhin\corref{cor1}}
\ead{Valentin.Irkhin@imp.uran.ru}
\cortext[cor1]{}
\address{Institute of Metal Physics, 620990, Ekaterinburg, S. Kovalevskaya str. 18, Russia}

\begin{abstract}
The quasistatic approximation and equation--of--motion decoupling for the electron Green's functions are applied to trace the effect of  electronic dispersion and electron correlations on the ferromagnetism of two--dimensional itinerant-electron systems. It is found that  next--nearest--neighbor hopping $t'$ is of crucial importance for ferromagnetism formation yielding the magnetic phase diagram which is strongly asymmetric with respect to half--filling. At small $t'$ in the vicinity of half--filling the ferromagnetic phase region is restricted by the spin--density wave instability,  and far from half--filling by one--particle (spin--polaron) instability. At $t'$ close to $t/2$ ferromagnetism is stabilized at moderate Hubbard $U$  due to substantial curvature of the Fermi surface which passes in the vicinity of the van Hove singularity points. The results obtained  are of possible importance for high--$T_{\rm c}$ compounds and layered ruthenates.
\end{abstract}

\end{frontmatter}

\section{Introduction}

During last two decades, the explanation of magnetic properties of itinerant--electron compounds attracts substantial interest in connection with physics of layered systems. The highly--correlated copper--oxide high--temperature superconductors, e.~g., La$_{2-x}$Sr$_x$CuO$_4$, demonstrate under doping both commensurate and incommensurate  antiferromagnetism \cite{Kastner:1998}, and the phase separation (PS) related to magnetic phase transitions \cite{Matsuda:2002}. Moderately correlated one-- and bilayered strontium ruthenates, Sr$_{2-x}$La$_x$RuO$_4$, Sr$_{2-x}$Ca$_x$RuO$_4$ and (Sr$_{1-x}$Ca$_x$)$_3$Ru$_2$O$_7$, demonstrate the interplay of (commensurate) ferro-- and (incommensurate) long--wave magnetic fluctuations \cite{Braden:2002}, and large Wilson ratio \cite{Qu:2008} indicating proximity to ferromagnetic instability, which is achieved under doping.
Explanation of drastically different magnetic properties of these two  compounds with  isomorphic crystal structure appears to be an intriguing challenge.

From a theoretical point of view, the physics of the layered compounds can be qualitatively captured within the one--band Hubbard model on the square lattice, basing on Cu or Ru $d_{xy}$--derived band and accounting for nearest ($t$) and next--nearest neighbor hopping ($t'$) integrals. ARPES and x--ray studies suggest the discrepancy between these compounds originating from substantial difference in their values of $t'$ and Coulomb (Hubbard) interaction $U$.

A large enough value of density of states (DOS) at the Fermi level (which is usually provided by van Hove singularity, VHS) leads to ferromagnetic ground state within the Stoner theory (mean--field theory of ferromagnetism). However, this approach neglects both Ne\'el antiferromagnetism and incommensurate magnetic ordering. Including these orderings into the mean--field theory \cite{Igoshev:2010} allows to trace the stability  of ground--state ferromagnetic order with respect to spin--density excitations depending on $t'$. It turns out that at small ratio $t'/t$ the ferromagnetic ordering is stable only at non--realistically large values of $U/t$ (even not too close to half--filling), and at smaller $U/t$ or close to half--filling the Ne\'el or spiral magnetic order is stable.

At larger $t'$ the van Hove singularity (i) provides the stability of the ferromagnetic state with respect to spin--flip one--particle excitations provided that the Fermi level lies in the vicinity of VHS, (ii) is well separated from the chemical potential position at half--filling where the most preferable magnetic orderings are characterized by large wave vector. It turns out that the competition of ferromagnetic and long--wave spin ordering in the vicinity of van Hove filling is a main limiting factor for ferromagnetic ordering. A priori it can be conjectured that the critical value of $U$ which is needed to stabilize ferromagnetic order is much smaller than in the case of small $t'$.

Another effect of finite $t'$ is strong dependence of PS region into the Ne\'el antiferromagnetic and spiral (at moderate $U$) or ferromagnetic (at large $U$) phases on the charge sign of current carriers \cite{Igoshev:2010}.

Above-discussed approximations completely miss correlation effects. The consideration of the case of large $U$ enables one to see the problem of ferromagnetism stability in new light.
Okabe \cite{Okabe:1998} applied a variational principle which provides some interpolation scheme between the cases of moderate and strong electronic correlations. It was found that at small number of current carriers (holes)  ferromagnetic ground state can be unstable with respect to spin--density wave formation, whereas at large hole number with respect to individual spin--flip. In Ref. \cite{Hanisch:1997} this problem was considered within a Gutzwiller--type method for  $t'\neq 0$. It was found that in the case of large $t'/t$ the ferromagnetic order acquires an additional stability for $n>1$ ($n$ is the electron concentration).

During last decade some weak-- and moderate coupling studies taking into account the correlation effects provided some progress in understanding  magnetic properties for non--zero $t'$. The spin--density wave instability of ferromagnetically ordered ground state was considered for rather large $t'$ ($t'\lesssim t/2$) within the quasistatic treatment of magnetic fluctuations (the quasistatic approximation) \cite{Igoshev:2007}. It was found that critical values of stability of ferromagnetic ordering substantially exceed those determined in the Stoner theory. The interplay of magnetic and electronic properties and their effect on the possibility of ferromagnetic instability was studied by the functional renormalization group (fRG) technique (see \cite{Igoshev:2011} and references therein).
Main result of these studies is that  small values $(t^{\prime }/t\leq0.2)$ favor the antiferromagnetic instability at Van Hove filling, moderate ones $(0.2\leq t^{\prime }/t\leq 0.35)$ the $d$--wave supeconducting instability, and  large $t^{\prime}/t>0.35$ correspond to ferromagnetic instability.

In the present paper we investigate the ferromagnetism problem for different  $t'$  within the quasistatic approximation and many-electron approach.

\section{Instabilities of ferromagnetic state: $t'/t$ far from $1/2$}
We start from the Hubbard model
\begin{equation}
\label{Hubbard_H}
\mathcal{H}=\sum_{\mathbf{k}\sigma}\epsilon_{\mathbf{k}}c^\dag_{\mathbf{k}\sigma}c_{\mathbf{k}\sigma}+U\sum_ic^\dag_{i\uparrow}c_{i\uparrow}c^\dag_{i\downarrow}c_{i\downarrow}.
\end{equation}
with the bare electronic spectrum $\epsilon_{\mathbf{k}}=-2t(\cos k_x+\cos k_y)+4t'(\cos k_x\cos k_y+1)$.

To treat the spin-wave instability of the ferromagnetic ground state we pass from the  model (1) to the spin--fermion model, which is justified in the vicinity of magnetically order state in two--dimensional systems \cite{Igoshev:2007}. We employ the quasistatic approximation for magnetic fluctuations in the effective field \cite{Igoshev:2007}. The static irreducible susceptibility reads
\begin{multline}
H_{\bf q}=\frac1{Z_{\xi\rightarrow\infty}}\int d^3\mathbf{S}\left[\frac13\Pi_{\parallel}(\mathbf{q}|S\, {\rm
sign}\,S^z)+\right.\\+\left.\frac23\Pi_{\perp}(\mathbf{q}|S\, {\rm sign}\,S^z)\right]\exp\left(-\mathcal{S}_{\xi\rightarrow\infty}\right),\label{G}
\end{multline}
where $Z_{\xi\rightarrow\infty}$ is a normalizing factor,
$$
\Pi_{\parallel,\perp}(\mathbf{q}|S)=-\frac1{N}\lim_{T\rightarrow0}\sum_{\mathbf{k}}\frac{f(\epsilon_{\mathbf{k}}-US)-f(\epsilon_{\mathbf{k}+\mathbf{q}}\mp US)}{\epsilon_{\mathbf{k}}-\epsilon_{\mathbf{k}+\mathbf{q}}-US\pm US}.
$$
$\mathcal{S}_{\xi\rightarrow\infty}=3U^2{\bf S}^2/(2\Delta^2)$ is gaussian spin fluctuation field action, specified by only one parameter $\Delta$ (having a sense of dispersion of the uniform static mode $\bf S$), which strongly simplifies the calculations.

Minimal critical value $U_{\rm c}$, which is necessary for the stability of ferromagnetic ordering, is determined by a generalized Stoner criterion
\begin{equation}
\label{gen_Stoner}
U_{\rm c}=1/H_{{\bf q}=0}.
\end{equation}
The r.h.s. of Eq. (\ref{gen_Stoner}) can be simplified via the relation
\begin{equation}
H_{{\bf q}=0}=\frac{1}{\Delta}\int\limits_{-4+8t'}^{4+8t'} \rho(\epsilon)
\varphi\left(\frac{\epsilon-\mu}{\Delta}\right)d\epsilon,
\label{Hi_irr1}
\end{equation}
where $\rho$ is bare DOS and $\varphi(x)=(3x^2+2)\exp(-{3 x^2}/2)/{\sqrt{6\pi}}$.

In fact, ferromagnetic state becomes stable when $H_{\bf q}$ as a function of $\bf q$ acquires a maximum at the point ${\bf q}=0$:
\begin{equation}
\label{DH>0}
\left[H_{\mathbf{q}=0}-H_{\mathbf{q}}\right]\geqslant0.
\end{equation}
The application of the criterion (\ref{DH>0}) together with Eq.(\ref{gen_Stoner}) allows to calculate $U_{\rm c}$ which determines the boundary of ferromagnetic and incommensurate magnetic phases  in the ground state at arbitrary $t'/t$.

Phase separation (PS) into N\'eel antiferromagnetic and ferromagnetic phases may also play an important role in our problem. Recently, the generalization of Visscher's method \cite{Visscher:1973} for PS boundary line  has been performed for the case of square lattice \cite{Igoshev:2010}:
\begin{equation}
\label{Visscher}
t/U_{\rm PS}(n)=[1+2{\rm sign}(1-n)t'/t]\pi(1-n)^2/2,
\end{equation}
One can see that PS boundary lines $n_{\rm PS}(U)$ possess strong asymmetry with respect to half--filling at finite $t'$.

In the limit of strong correlations we perform also calculations of the electron Green's functions using the equation-of-motion method in the many--electron representation of Hubbard's X--operators \cite{Irkhin}. Note that within this approach we  treat only the destruction of saturated or non--saturated ferromagnetic ground state by one-particle (individual) spin--flip excitations.  In the simplest Hubbard--I approximation the Green's functions read
\begin{equation}
G_{\mathbf{k}\sigma }(E)=[F_{\sigma }^0(E)-\varepsilon _{\mathbf{k}}]^{-1},
 \label{eq:G}
\end{equation}
where the bare inverse locator $F_{\sigma }^0(E)$ has a form
\begin{equation}
F_{\sigma }^{0}(E)=\frac{E(E-U)}{E-U(n_{0}+n_{\sigma })},
 \label{eq:F}
\end{equation}
$n_\alpha $ being the number of holes ($\alpha =0$) or singly-occupied states with spin projection $\sigma $. However, this approximation is inapplicable to the ferromagnetism problem since it violates important kinematic relations. It was improved in Ref.~\cite{Irkhin} by including spin and charge fluctuation to obtain  $F_{\sigma }^{0}(E)\to F_{\mathbf{k}\sigma }(E)$, where $F_{\mathbf{k}\sigma }$ is the corrected inverse locator which is determined self-consistently.
This many-electron (ME) approximation  allows to  calculate the quantities $n_{\sigma }$ and obtain the equation for magnetization. Solving this equation yields two magnetic phase transitions: from saturated ferromagnetism (sFM) to non--saturated ferromagnetism (nsFM), and from non-saturated ferromagnetism to paramagnetic state (PM). The first instability has in fact spin-polaron nature (occurrence of the pole of the spin-down Green's function below the Fermi level).

\begin{figure}[htb]
\begin{center}
\includegraphics[width=0.73\columnwidth,angle=-90]{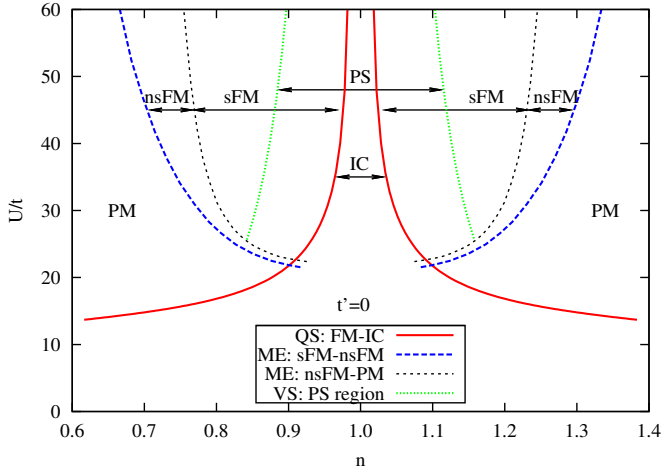}
\end{center}
\caption{Ground state magnetic phase diagram in variables $(n,U)$. $t'=0$. Comparison of different types of instability of ferromagnetic ground state. ``QS: FM--IC'' is boundary line between ferromagnetic and incommensurate magnetic phase regions obtained within quasistatic approach. ``ME: sFM--uFM'' is the boundary line between saturated and unsaturated ferromagnetic phase regions obtained within ME approach, ``ME: uFM--PM'' is boundary line between  unsaturated ferromagnetic and paramagnetic phase regions obtained within ME approach \cite{Irkhin}, ``VS: PS region'' is boundary region of PS to saturated ferro-- and antiferromagnetic phases obtained within Visscher method (see \cite{Igoshev:2010})}
\label{fig:00}
\end{figure}

\begin{figure}[htb]
\begin{center}
\includegraphics[width=0.73\columnwidth,angle=-90]{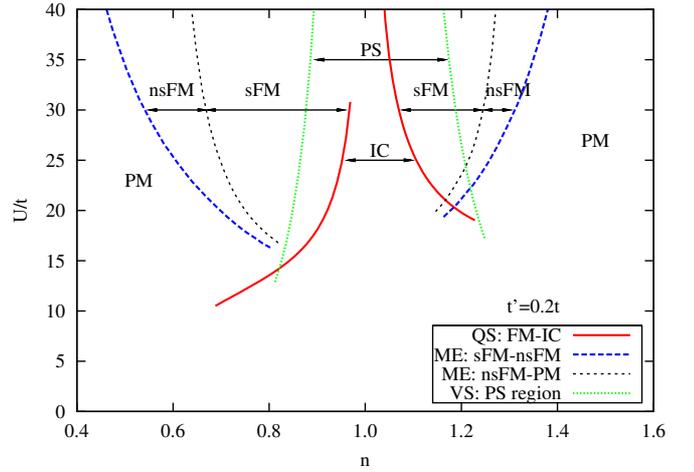}
\end{center}
\caption{The same as in Fig. \ref{fig:00}, $t'=0.2t$}
\label{fig:20}
\end{figure}

\begin{figure}[htb]
\begin{center}
\includegraphics[width=0.73\columnwidth,angle=-90]{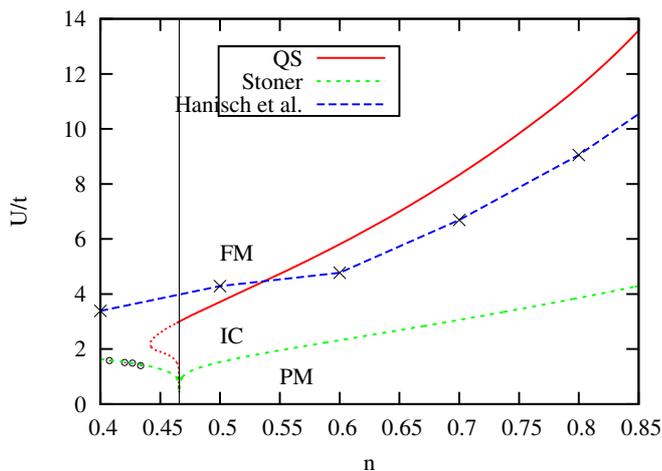}
\end{center}
\caption{Ground state magnetic phase diagram in variables $(n,U)$. $t'=0.45$. Comparison of results obtained within quasistatic approach (``QS''), Stoner theory (``Stoner'') and Gutzwiller--type approach (``Hanisch'') \cite{Hanisch:1997} is shown}
\label{fig:45}
\end{figure}

The phase diagrams calculated by two methods described are shown at small $t'/t$ in Figs. \ref{fig:00} and \ref{fig:20}. We see that the spin--density wave  (Eqs. (\ref{gen_Stoner}) and (\ref{DH>0})) and  one--particle (Eqs. (\ref{eq:G}) and (\ref{eq:F})) instability lines restrict the ferromagnetic region from different sides.

The ferromagnetic state for  $t'=0$ occurs for very large (practically unrealistic) $U$ only because of competition with commensurate or incommensurate antiferromagnetism. With increasing $t'$, the van Hove singularity  (VHS) in the electron spectrum, which determines ferromagnetic instability, is shifted from the band centre, and the critical $U$ considerably decreases for $n<1$. Thus the next--nearest neighbor hopping is of crucial importance for ferromagnetism, the magnetic phase diagram being strongly asymmetric with respect to half-filling.

\section{Instabilities of ferromagnetic state: $t'/t$  close to $1/2$}
In the case of large $t'$ (in our calculations, $t'=0.45t$) ferromagnetism is practically absent (at not too large $U/t$) for $n>1$. Hence we focus our attention on the case $n<1$. When the Fermi level lies in the vicinity of VHS, two opposite tendencies take place: (i) the stability of ferromagnetic order
substantially increases due to large DOS at the Fermi level (critical $U$ decreases), (ii) long--wave spin-density instability (rather than the N\'eel antiferromagnetism) becomes actual ferromagnetism's competitor, as opposed to the case of small $t'/t$ (critical $U$ increases).

Since critical $U$ necessary for the ferromagnetism stability is not large, we do not apply the strong--coupling ME approach used in previous Section.
The results of the quasistatic approximation \cite{Igoshev:2007} are shown in Fig. \ref{fig:45}. A comparison with the results of Ref. \cite{Hanisch:1997} at $t'\approx0.43t$ is also shown. The phase diagram is strongly asymmetric with respect to VHS: when Fermi level lies above VHS,  competition with long--wave magnetic order increases critical $U$. At the same time, when Fermi level lies below VHS, ferromagnetic ordering is stabilized by rather small Coulomb interaction. The asymmetry is connected with the electronic topological transition which occurs when the Fermi level crosses VHS.
At the Fermi level above VHS,  flat parts of the Fermi surface occur in the vicinity of VHS points (``quasinesting'' situation), which results in enhancement of the incommensurate fluctuations. This conclusion is confirmed by the functional renormalization group studies \cite{Igoshev:2011}. We can also see that our results are in qualitative agreement with the results of the Gutzwiller--type method \cite{Hanisch:1997}. However, the latter does not capture delicate features of incommensurate fluctuations properties, in particular, strong dependence on the Fermi level position.

To conclude, we calculated the boundaries of ferromagnetic region on the ground state phase diagram at different $t'$. We found that ferromagnetic region is restricted by instabilities of two types: spin--wave instability and single--particle instability. It was shown that the Stoner criterion is not applicable for both small and large $t'$. Indeed, at small $t'$ in the vicinity of half--filling the ferromagnetism can be destroyed by collective spin--wave excitations with large wave vector. Away of half--filling the ferromagnetic order is destroyed by single--particle excitations, but such an (spin--polaron) excitation causes strong many--electron renormalizations and should be treated with account of  electron correlations. At large $t'/t\lesssim1/2$ the unique possibility of ferromagnetic order occurs in the vicinity of van Hove singularity at relatively small $U$, but the long--wave spin--density instability substantially suppresses the ferromagnetic order in comparison with the Stoner criterion.

The above results for the case of large $t'/t$ yield a possibility to explain the magnetic properties of doped ruthenates like Sr$_{2-x}$La$_x$RuO$_4$, where the ferromagnetic order appears to be suppressed  \cite{Kikugawa:2004} despite  that Fermi level is close to VHS of $d_{xy}$--derived band and the Fermi energy is above VHS \cite{Shen:2007}. 
Since tight--binding fit for the latter band yields $t'/t\sim 0.4$ \cite{Monthoux:2005} and x--ray studies estimate $U$ as about $3.5t$ \cite{Yokoya:1996}, we can explain the ferromagnetism suppression by competition with incommensurate long--wave magnetic order. Moreover, one can expect that hypothetical increase of available range of La concentration will induce the transition into the ferromagnetic state at the $x$ value where the Fermi energy coincides with VHS of $d_{xy}$--derived band. Thus the difference in electron spectrum of layered systems (e.g., of cuprates and ruthenates) strongly influences their magnetism.



This work was supported in part by  the project Quantum Physics of Condensed Matter of the Presidium of RAS, President Program of Scientific Schools Support 4711.2010.2, Partnership Program of the Max–Planck Society, ``Dynasty'' foundation, projects of RFBR No. 11-02-00931-a, No. 11-02-00937-a, and the project of young scientists of Ural Branch of RAS M--8.

\end{document}